\documentclass[twocolumn]{aastex7}

\usepackage{microtype}
\usepackage{url}
\usepackage{amsmath}
\usepackage{mathtools}
\usepackage{esint}
\usepackage{amssymb}
\usepackage{multirow}
\usepackage{graphicx}
\usepackage{scalerel}
\usepackage{calc}
\usepackage{etoolbox}
\usepackage{marginnote}
\usepackage{nicefrac}
\usepackage{tabstackengine}
\usepackage{diagbox}
\usepackage[makeroom]{cancel}
\usepackage{mathdots}
\usepackage{bbm}
\usepackage{booktabs}
\usepackage{xspace}
\usepackage{upgreek}
\usepackage[T1]{fontenc} 
\usepackage{textcomp}
\usepackage{ifxetex}
\ifxetex
\usepackage{fontspec}
\defaultfontfeatures{Extension = .otf}
\fi
\usepackage{fontawesome}
\usepackage{listings}
\usepackage{mathtools}
\stackMath

\newcommand{\pycodelink}[1]{\href{#1}{\codeicon}\@\xspace}
\newcommand{\jlcodelink}[1]{\href{#1}{\codeicon}\@\xspace}
\newcommand{\animlink}[1]{\href{#1}{\animicon}\@\xspace}
\newcommand{\prooflink}[1]{\href{#1}{\raisebox{-0.1em}{\prooficon}}}

\renewcommand{\eqref}[1]{\ref{eq:#1}}

\definecolor{linkcolor}{rgb}{0.1216,0.4667,0.7059}
\newcommand{\codeicon}{{\color{linkcolor}\faFileCodeO}}
\newcommand{\prooficon}{{\color{linkcolor}\faPencilSquareO}}
\newcommand{\animicon}{{\color{linkcolor}\faPlayCircle}}

\newtagform{eqtag}[]{(}{)}
\newcommand{\currentlabel}{None}

\newenvironment{proof*}[1]{%
\ifstrempty{#1}{%
\renewtagform{eqtag}[]{\raisebox{-0.1em}{{\color{red}\faPencilSquareO}}\,(}{)}%
}{%
\renewtagform{eqtag}[]{\prooflink{#1}\,(}{)}%
}%
\usetagform{eqtag}%
\renewcommand{\currentlabel}{#1}
\equation%
}{%
\endequation%
\renewtagform{eqtag}[]{(}{)}%
\usetagform{eqtag}%
\message{<<<\currentlabel: \theequation>>>}
}

\newcommand{\T}{\ensuremath{\mathrm{T}}}

\newcommand{\x}{\ensuremath{\mbox{$x$}}}
\newcommand{\y}{\ensuremath{\mbox{$y$}}}
\newcommand{\z}{\ensuremath{\mbox{$z$}}}

\DeclareMathAlphabet\mathbfcal{OMS}{cmsy}{b}{n}

\definecolor{dim}{rgb}{0.8,0.8,0.8}
\usepackage{listings}
\definecolor{codegreen}{rgb}{0,0.6,0}
\definecolor{codegray}{rgb}{0.5,0.5,0.5}
\definecolor{codepurple}{rgb}{0.58,0,0.82}
\definecolor{backcolour}{rgb}{0.95,0.95,0.95}
\lstdefinestyle{mystyle}{
    backgroundcolor=\color{backcolour},
    commentstyle=\color{codegreen},
    keywordstyle=\color{magenta},
    numberstyle=\tiny\color{codegray},
    stringstyle=\color{codepurple},
    basicstyle=\small\ttfamily,
    breakatwhitespace=false,
    breaklines=true,
    captionpos=b,
    keepspaces=true,
    numbers=left,
    numbersep=5pt,
    showspaces=false,
    showstringspaces=false,
    showtabs=false,
    tabsize=2,
    aboveskip=1em,
    belowskip=1em,
    keywords=[2]{map},
    keywordstyle=[2]{\color{black!80!black}},
}
\newcommand{\jorbit}{\texttt{jorbit}\@\xspace}

\newcommand{\lightkurve}{\texttt{lightkurve}\@\xspace}

\accepted{January 4, 2026}
\submitjournal{The Astronomical Journal}
\shorttitle{Minor Planets in TESS TPFs}
\shortauthors{Cassese et al.}
\graphicspath{{./}{figures/}}

\usepackage{amssymb}
\usepackage{bm}
\usepackage{wrapfig}

\begin{document}

\title{Quantifying the Contamination of TESS Ecliptic-Plane Light Curves by Minor Planets}


\author[0000-0002-9544-0118]{Ben Cassese}
\email{b.c.cassese@columbia.edu}
\affiliation{Dept. of Astronomy, Columbia University, 550 W 120th Street, New York NY 10027, USA}

\author[0000-0003-1481-8076]{Justin Vega}
\email{j.vega@columbia.edu}
\affiliation{Dept. of Astronomy, Columbia University, 550 W 120th Street, New York NY 10027, USA}

\author[0000-0003-4755-584X]{Daniel A. Yahalomi}
\altaffiliation{LSST-DA DSFP Fellow}
\email{daniel.yahalomi@columbia.edu}
\affiliation{Dept. of Astronomy, Columbia University, 550 W 120th Street, New York NY 10027, USA}

\author{David Gelpi}
\altaffiliation{Columbia STAR Program Student}
\email{dgelpi2026@columbiasecondary.org}
\affiliation{Columbia Secondary High School, 425 W 123rd Street, New York NY 10027, USA}

\author{Eva Marmolejos}
\altaffiliation{Columbia STAR Program Student}
\email{emarmolejos2027@columbiasecondary.org}
\affiliation{Columbia Secondary High School, 425 W 123rd Street, New York NY 10027, USA}

\author{Aneisa Rampersaud}
\altaffiliation{Columbia STAR Program Student}
\email{arampersaud2027@columbiasecondary.org}
\affiliation{Columbia Secondary High School, 425 W 123rd Street, New York NY 10027, USA}

\author{Aware Deshmukh}
\email{and2161@columbia.edu}
\affiliation{Dept. of Astronomy, Columbia University, 550 W 120th Street, New York NY 10027, USA}

\author[0000-0003-4540-5661]{Ruth Angus}
\email{rangus@amnh.org}
\affiliation{Department of Astrophysics, American Museum of Natural History, 200 Central Park West, Manhattan, NY, USA}

\author[0000-0002-7670-670X]{Malena Rice}
\email{malena.rice@yale.edu}
\affiliation{Dept. of Astronomy, Yale University, New Haven, CT 06511, USA}

\begin{abstract}
Though missions devoted to time series photometry focus primarily on targets far beyond the solar system, their observations can be contaminated by foreground minor planets, especially near the ecliptic plane where solar system objects are most prevalent. Crucially, depending on one's choice of data reduction/background estimation algorithm, these objects can induce both apparent brightening and/or dimming events in processed light curves. To quantify the impact of these objects on archived TESS light curves, we used N-body integrations of all currently known minor planets to postdict all 600,000+ of their interactions with stars selected for high-cadence observations during TESS ecliptic plane sectors. We then created mock images of these moving sources and performed simple aperture photometry using the same target and background apertures used in SPOC processing. Our resulting 10,000+ target-specific light curves, which faithfully model the time-dependent positions and magnitudes of the actual solar system objects that approached each target, reveal that $>95\%$ of high-cadence ecliptic plane targets experience a minor planet crossing within 1 TESS pixel of the source. Additionally, 50\% of all $T>13$\,mag targets experience at least one instantaneous moment where the contaminating flux from minor planets exceeds 1\% of the target flux. We discuss these population-level results and others, and highlight several case studies of bright flybys.
\end{abstract}

\keywords{Asteroids (72) --- Transit photometry (1709) --- Time series analysis (1916)}

\section{Introduction} \label{sec:intro}

In recent years several space-based missions including, among others, \textit{Kepler}/\textit{K2} \citep{borucki_2010, howell_2014} and the Transiting Exoplanet Survey Satellite  \citep[TESS;][]{ricker_2015} have revolutionized subfields of study built on time series photometry. Although both missions were designed primarily to hunt for planets in other star systems, the pair has enabled a wide array of analyses in transient phenomena more generally \citep{winn_2024}.

Over the course of their missions, both of these spacecraft spent portions of their pointings aimed at fields located near the ecliptic plane. In the case of \textit{K2}, this was by necessity following mechanical failures aboard \textit{Kepler}, while in the case of TESS, this was by choice, one made in part to enable follow up of \textit{K2} discoveries. This particular band of the sky distinguishes itself from other fields by hosting the vast majority of solar system objects, or SSOs. These objects, which are almost always unresolvable point sources, can be identified by their motion in the plane of the sky relative to background stars.

Some investigators focus on these objects themselves, and indeed, TESS especially has enabled numerous studies of asteroid rotation period measurements and shape reconstruction \citep[e.g.][]{pal_2020, woods_2021, mcneill_2023, takacs_2025, vavilov_2025}. Its enormous field of view and ability to collect uninterrupted time series render it a unique facility capable of unlocking novel regimes of asteroid dynamics. It is expected to continue contributions to solar system science in years to come, both through analyses of new observations and processing of its rich archives \citep[e.g.][]{rice_2020}.

However, solar system science accounts for only a fraction of the total scientific output of TESS observers, and the majority of TESS-based studies focus on much more distant objects whose positions are effectively fixed on the sky. Thus, changes in flux surrounding a particular position are usually attributed entirely to the objects known to always be near that position. When observing in the ecliptic plane, this simplification overlooks the contributions of SSOs which can pass into and out of a given frame over the course of a time series.

These moments of coincident nearby sky-plane alignment could, depending on the exact data processing steps, cause the source to appear temporarily brighter \textit{or fainter} depending on an SSO's placement relative to the extraction and background apertures. Consequently, SSOs can, in principle, mimic many different transient events, from stellar flares to exoplanet transits. There is a great potential benefit in quantifying the typical amplitude, frequency, and timescale of this contamination to assess the risks to other science cases.

Thankfully, centuries of asteroid observations have already uncovered the majority of SSOs bright enough to meaningfully affect most TESS observations \citep[see, e.g.,][]{hendler_2020}, and in theory it should be possible to pre or postdict the contributions of SSOs on a given TESS dataset. However, until recently, this forward modeling was quite challenging to do with local computing resources. While it is straightforward to take a given SSO and derive an ephemeris via a dynamical model of the solar system \citep[e.g. JPL DE440,][]{park_2021}, the \textit{inverse} problem is more demanding. Given a patch of sky and a time, determining which SSOs are nearby requires propagating \textit{all} known SSOs to that time, evaluating their positions, then repeating this exercise for every time in the time series. Most often, those interested in identifying nearby SSOs at a single specific time had to rely on external tools like the Minor Planet Center's ``MPChecker'' service\footnote{\href{https://www.minorplanetcenter.net/cgi-bin/checkmp.cgi}{MPChecker Web Service}} or the Virtual Observatory of the IMCEE CNRS\footnote{\href{https://ssp.imcce.fr/webservices/skybot/}{SkyBoT: The Virtual Observatory Sky Body Tracker}}, then string many of these queries together into a time series.

The \jorbit software package \citep{jorbit}, described further in Sec. \ref{sub:jorbit}, was designed in part to tackle this challenge and enable local identification and simulation of all minor planets (SSOs that are neither planets nor comets) near a given coordinate at given times. In this study, we use \jorbit to quantify the effects of minor planets on TESS observations by:

\begin{enumerate}
    \item Identifying all of the known minor planets that may have affected every star selected for high-cadence 20\,s observations in sectors 42-46 and 70-72 (all of the pre-2025 ecliptic plane observations). In principle this analysis could be repeated for any TESS target regardless if it is extracted from 20\,s, 2\,min, or Full Frame Image data, and our limited focus on the 20\,s cadence observations was only to bound the scope of this study.
    \item For each star/sector combination, creating a model Simple Aperture Photometry light curve of minor planet contributions to the total flux. These are built on accurate minor planet trajectories, realistic time-varying minor planet magnitudes, a model Pixel Response Function, and target-specific background and extraction apertures.
    \item Analyzing the collection of simulated light curves to extract population-level distributions, and highlighting several extreme case studies.
\end{enumerate}

Throughout this article, we use ``target'' to refer to observations of a specific source from the TESS Input Catalog (TIC) in a specific sector, as many of the target stars were observed in multiple sectors. In total, we investigated 10,610 unique star/sector Target Pixel Files (TPFs) of 5,543 unique target stars\footnote{In Sector 72, TIC 184842717, a bright star selected by GI program G06200, does not have an aperture or background mask associated with its Target Pixel File. This is noted in the Data Release Notes for Sector 72. We consequently excluded this single target from our analysis.}. Additionally, we use the target-specific background and extraction apertures (clusters of potentially non-contiguous pixels within each TPF) automatically assigned by the Science Processing Operations Center (SPOC) pipeline hosted at NASA Ames \citep{jenkins_2016}. Many other high-level science products derived from data transmitted from the TESS spacecraft, such as products from the TESS quick-look pipeline  \citep[QLP;][]{huang_2020}, \texttt{eleanor} \citep{feinstein_2019}, TESS-Gaia Light Curves \citep{han_2023}, and T16 \citep{hartman_2025}), are stored on the Mikulski Archive for Space Telescopes where they can also be retrieved using community tools such as \lightkurve \citep{lightkurve}. However, we only attempted to emulate the SPOC Simple Aperture Photometry (SAP) procedure with our mock image cubes, and therefore our results are not directly comparable to products that underwent more advanced correction routines.

We describe our postdiction simulation process in Sec. \ref{sec:simulations}, population-level results in Sec. \ref{sec:general_results}, a handful of interesting individual case studies in Sec. \ref{sec:case_studies}, and conclude in Sec. \ref{sec:conclusion}. While we do not plan to distribute our entire $>2$\,TB simulated dataset due to storage difficulties, we encourage those interested in bespoke simulations for individual targets to contact the authors.

\section{Postdiction Simulations} \label{sec:simulations}

We identified all minor planet-TPF crossings through a two-step process: an initial simulation that identified all potentially relevant pairs, and a post-processing step that created artificial light curves from the modeled trajectories. Each step is described in its own subsection below.

\subsection{\texttt{jorbit}'s Algorithms} \label{sub:jorbit}

We used \jorbit \citep{jorbit}, a newly-developed minor planet simulation framework, to postdict the impact of foreground minor planets on pre-2025 TESS ecliptic plane light curves for targets observed at 20\,s cadence. Here we provide a brief summary of \jorbit's general underlying algorithms before describing our specific application, though we note that a more complete description of the software can be found in \citet{jorbit}.

\jorbit's \texttt{nearest\_asteroid} function relies on a cached and compressed simulation of 1,438,635 minor planets\footnote{Not comets: \jorbit will fail to flag any cometary material as it did not include any in its cached simulations. In general, comets must be handled differently than minor planets, as their outgassing produces strong non-gravitational accelerations that violate both Keplerian motion and the assumptions of more complex N-body simulations like JPL Horizons. The coefficients describing these accelerations are specific to each comet, and it is challenging to integrate them forward in time for more than a couple of years if that time span includes their perihelion. Note, however, that there are only 4599 currently known comets in the Minor Planet Center archives compared to the $>1.4$ million minor planets, and thus their exclusion is not expected to impact our results.} known as of early February 2025. Each of these objects, whose names and absolute magnitude $H$ were retrieved from the Minor Planet Center's MPCORB.DAT file, were assigned an initial condition based on JPL Horizons'\footnote{\href{https://ssd.jpl.nasa.gov/horizons/}{JPL Horizons Web Service}} estimate of their state vectors on 2020-01-01. Each object was then integrated forwards and backwards in time for 20 years under the influence of Newtonian gravity from the sun and perturbations from 26 massive objects (the planets, the moon, and the large asteroids included in the standard DE440 ephemeris). The locations of these perturbers were computed via local parsing of the JPL DE440 ephemeris \citep{park_2021} at every time step following the standard practice of other high-precision solar system integrators \citep{holman_assist_2023}. The 3D positions of the minor planets were logged every 10 hours and converted to sky-projected coordinates as seen from the geocenter while accounting for particle-specific light travel times. The RA and Dec of these coordinates were then separately compressed as piecewise 11th-order Chebyshev polynomials broken into 30-day chunks, mimicking the format of the DE440 ephemeris.

When a user queries \jorbit for the identities and trajectories of all minor planets that approached within a given radius of a given coordinate during a given time span, \jorbit first uses the appropriate chunk of polynomial coefficients to compute the on-sky position of every minor planet at the midpoint of the time series. It then drops all objects that were $>30^\circ$ from the target at that time to avoid additional computation spent on particles with little chance of reaching the region of interest. Then, for every step in the time series, it computes the positions of all remaining minor planets on this culled list and flags any that fell within the specified radius of the target, as seen from the geocenter (recall that the polynomials it is evaluating were based on a simulation that assumed the observer was at the geocenter).

Finally, for every minor planet on this most selective list, \jorbit recalls a cached estimate of their 3D barycentric position/velocity at epoch 2020-01-01 as computed by JPL Horizons in February 2025, then uses these as the initial conditions in a new N-body simulation to recover the particles' underlying 3D trajectories. This simulation uses a user-specified acceleration function, meaning its accuracy depends on how many perturbative effects the user wishes to include (e.g. how many planets, whether to use Newtonian or general relativistic gravity, or whether to include gravitational harmonics). This integration is necessary since we require the 3D position to accurately compute the effects of parallax and phase-dependent magnitude. Finally, \jorbit queries JPL Horizons for the barycentric position of the observer at each time step, then combines this with the particle trajectories to compute the on-sky positions as seen from the observer's true (non-geocentric) position, as well as each particle's time/phase-angle specific apparent visual magnitude.

The runtime of this process depends on the length and resolution of the time series in question, the density of minor planets near the requested coordinate, and the exact settings of the acceleration function used. Anecdotally, it took a few minutes to generate a sector-long, 1\,min cadence model light curve for each of the TESS targets investigated in this study. In hindsight, the runtime could have been dramatically improved by using a coarser cadence. These computations could be carried out on a laptop, but we used the Grace high-performance computing cluster at Yale University and the Anvil cluster \citep{anvil} at Purdue University with an allocation granted via the NSF ACCESS program \citep{nsf_access} to carry out the bulk of the simulations. Those interested in replicating this analysis for specific targets rather than a bulk of TESS data can expect to run their computations locally.

\subsection{Specific Application to TESS} \label{sub:jorbit_tess_application}

Above we described \jorbit's general approach to identifying and propagating all minor planets that approach a given coordinate during a given time span. Here we will discuss how we applied this tool to quantify the impact of minor planet crossings on TESS light curves and diagnose potential shortcomings to this procedure, and we will outline our approaches to mitigate each one where appropriate. We identify four choices that in principle could result in a minor planet incorrectly not being identified:

\begin{enumerate}
    \item The minor planet/comet was not included in \jorbit's cache
    \item The object was moving quickly enough to be cut during \jorbit's down selection
    \item The use of a simple acceleration model (i.e. Newtonian gravity only) led to a significantly incorrect particle trajectory
    \item A particle falls outside the radius of interest when viewed from the geocenter, but within it when viewed from TESS due to parallax
\end{enumerate}

We cannot fully mitigate the first risk: if a minor planet was not included in the MPCORB.DAT file in early February 2025, either because it was a comet or not yet discovered, \jorbit does not know about it and consequently will never identify it. Note that this cutoff applies to this particular work only, since \jorbit will periodically update its cached simulations to maintain agreement with the MPC and JPL Horizons. 

The second risk is unlikely to meaningfully affect the total flux in our final model but could potentially lead to the exclusion of rapidly-moving Near Earth Objects (NEOs). Since \jorbit makes an initial cut in the minor planets it considers based on the location of every object at the midpoint of the time series, we will miss any particles that were $>30^\circ$ from the target in the middle of the sector but were moving fast enough that they fell into our radius of interest at some other time. The $\sim28$\,day length of TESS sectors considered here implies that the particle would have to sustain a $>2^\circ$/day sky velocity for many days, a rate comparable to NEOs moving at their fastest. We accept this tradeoff between completeness and computational speed and note that most NEOs are faint enough to not meaningfully affect our conclusions even when left out. We do wish to highlight, however, that there is a risk to running single \jorbit queries over very long time spans, where ``very long'' depends on the sky-plane velocity of the particles of interest. In the context of TESS observations, for multi-sector or extended-stare studies (e.g., target stars in the Continuous Viewing Zone or Sectors 97 and 98), one should run \jorbit's \texttt{nearest\_asteroid} function on several shorter time spans, which each would have their own 30-degree-at-midpoint cutoff.

The third potential pitfall is similarly unlikely to affect our model. This was the choice to use Newtonian rather than relativistic gravity in the two stages of N-body simulations (first \jorbit's large initial simulation, then in the target-specific follow up with just the particles of interest). Though \jorbit is capable of considering Parameterized Post-Newtonian gravity based on \citet{newhall_1983} and the implementation in \texttt{REBOUNDx} \citep{reboundx}, we chose to use the simpler model for computational considerations. Considering that relativistic corrections typically shift main belt positions $<1$\arcsec\, on the timescales considered here, it is improbable that neglecting these relativistic corrections would nudge a particle out of the region of interest. Note that we still included perturbations from the planets and several large asteroids, which are important over the $\sim$years timescale considered here (from initial conditions on 2020-01-01 to each TESS sector).

The fourth potential pitfall, however, demands greater care to address. All of \jorbit's cuts/decisions about which minor planet to integrate are based on the cached simulation that placed the observer at the geocenter. However, since no observer will actually collect data from this location, we must consider the parallax introduced by shifting the observer to their true position. For any ground-based observer looking at a main belt asteroid, the parallax will be comparable to 10\arcsec. For a several arcmin field of view, this implies that we can inflate our search radius by just a few percent to avoid the possibility of a target falling outside the search cone from the perspective of the geocenter, but within it from the perspective of the true observer. This introduces a new problem, that a minor planet appears in the search cone from the perspective of the geocenter but \textit{not} from the perspective of the observer. But, filtering these superfluous cases introduces much less overhead than initially simulating them.

TESS, however, is not on the Earth's surface, or indeed even in low Earth orbit. The spacecraft occupies an eccentric orbit with a $>60$\,R$_\oplus$ apogee and a period that places it within the 2:1 resonance with the Moon to average out lunar perturbations \citep{gangestad_2013}. This is a much larger lever arm than just one Earth radius, and, consequently, the parallax will be much larger.

\begin{figure}
    \centering
    \includegraphics[width=\columnwidth]{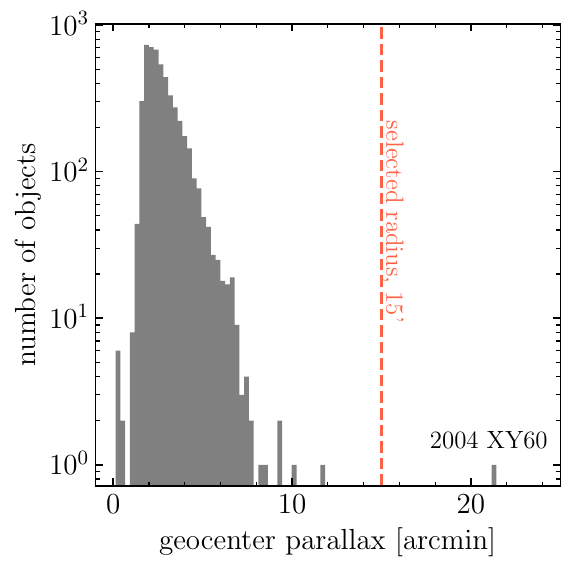}
    \caption{The distribution of differences in sky position for randomly sampled minor planets as seen from TESS and the geocenter in January 2025. As described in Sec \ref{sub:jorbit_tess_application}, we used this biased distribution to set our search radius. The outlier NEO (289227) 2004 XY60 was the only randomly sampled object with a parallax $>15\arcmin$.}
    \label{fig:tess_parallax}
\end{figure}

To quantify an expected range of parallaxes to determine the factor by which we need to inflate our search radius, we first randomly sampled 5,000 numbered minor planets and queried JPL Horizons for their astrometric coordinates as seen from the geocenter and from TESS once a day between 2025-01-01 and 2025-01-31. We then recorded the maximum difference between the two positions during this time span for each object and used this as an estimate of the worst-case parallax. This is an imperfect test, since a large fraction of randomly selected objects will not be observable during these times: they will be on the opposite side of the sun, where they are further away than their visible counterparts, and thus have the effects of parallax suppressed. Our resulting distribution of parallaxes, shown in Fig. \ref{fig:tess_parallax}, is thus biased towards smaller values than we would typically see for observable objects.

Though biased, this distribution does allow us to heuristically pick a search radius wide enough to nearly ensure that any object that passed near enough to a target star to affect its 3.85\arcmin x3.85\arcmin\footnote{This is the typical size of an 11x11 pixel TPF, though certain targets may have different sizes/aspect ratios, especially if the target is expected to saturate the detector, e.g. TIC 350347140 which is 29x11 pixels.} TPF would also be caught in the geocentric search cone. Though one object, the NEO (289227) 2004 XY60, would experience a $>20'$ parallax if viewed from the geocenter vs. from TESS during January 2025, all others shifted by $<15'$. Accordingly, we ran all \jorbit queries with a search radius of 15\arcmin, a much larger radius than the actual target of interest, and used the post-processing step to thin our final list of objects to those which actually approached each TPF. We acknowledge, however, that a small fraction of exceptionally close objects might still be missed by this approach, similar to our $30^\circ$ cut described above.

\subsection{Post processing and model light curve generation} \label{sub:postprocessing}

After running \texttt{nearest\_asteroid} with the setting described above on the Grace high-performance computing cluster at Yale University, we were left with $>2$\,TB of intermediate data products. These largely consisted of ephemerides of minor planets that did not contribute to the flux of their TPFs thanks to the search radius inflation described in Sec. \ref{sub:jorbit_tess_application}. To extract meaningful information from these simulations, we compressed each into three products: a list of all minor planets that contributed the equivalent of at least a $T=24$\,mag point source in either the extraction or background aperture at any point in the time series; a time series of the total flux from minor planets in the SPOC extraction aperture; and a time series of the total flux from minor planets in the SPOC background aperture.

To create these processed products we carried out the procedure illustrated in Fig. \ref{fig:postprocess_procedure} and described below:

\begin{figure*}
    \centering
    \includegraphics[width=0.8\textwidth]{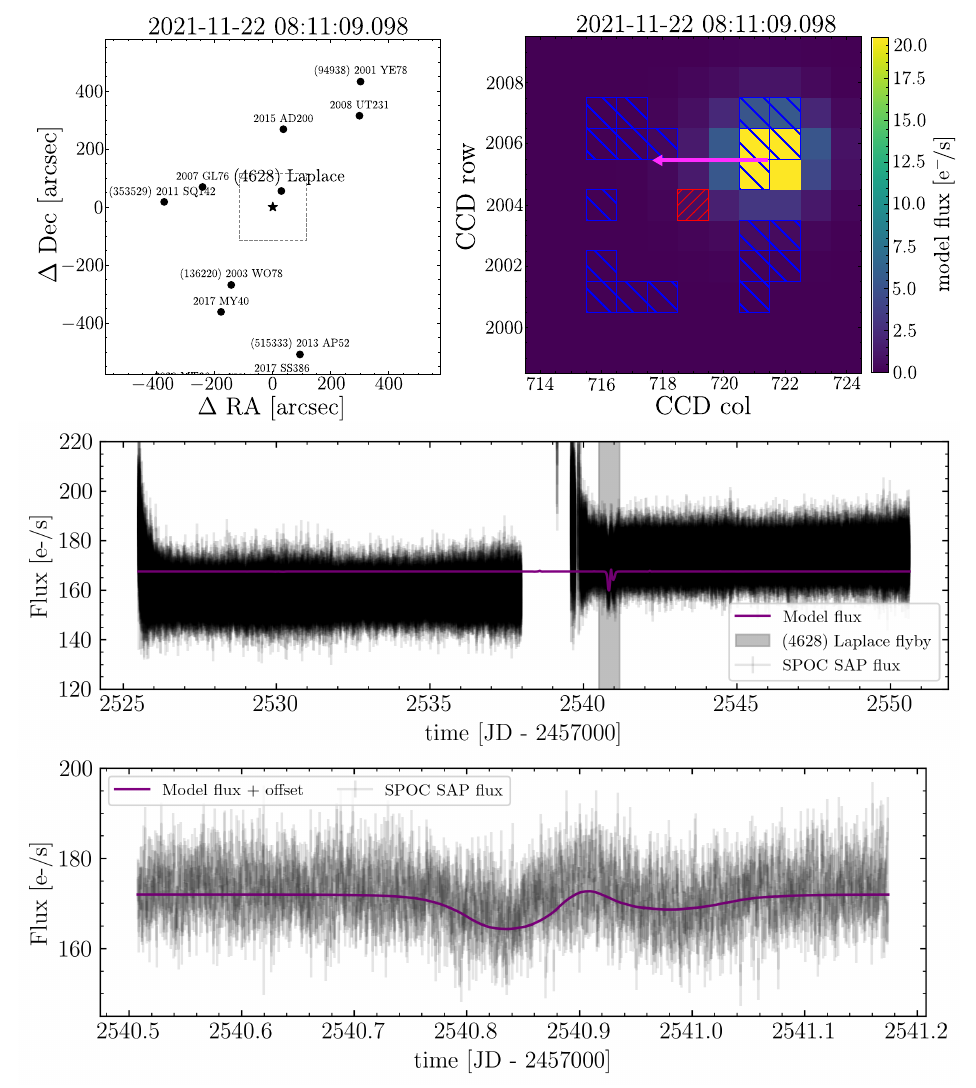}
    \caption{An illustration of our modeling procedure for TIC 81720355 ($T=14.88$\,mag), Sector 45. Top left: \jorbit's predictions for locations of all minor planets surrounding the target at one specific time. The dotted square has a side length of 3.85\arcmin, the width of the TPF. Throughout the sector, \jorbit flagged and created ephemerides for 174 minor planets that passed within 15\arcmin\, of the target as seen from the geocenter. The postprocessing step revealed that only 42 of these minor planets, including (4628) Laplace ($H=11.27$, $T=14.87$\,mag at closest approach), actually contributed flux to the extraction or background apertures set by SPOC. Top right: the conversion from the predicted minor planet locations to a model image. The SPOC target extraction aperture is marked with the red hatched pixel, while the SPOC background aperture is marked with the blue hatched pixels. The pink arrow shows the approximate instantaneous on-sky velocity vector. Note at this time slice (4628) Laplace contributes to the background aperture but not the extraction aperture, and that it is headed for another cluster of background pixels. This panel time corresponds to the moment of lowest predicted model flux in the lower two panels. Middle: the full Sector 45 SAP light curve for TIC 81720355 (retrieved via \lightkurve) and our model, overplotted. Note that this is a prediction, not a fit: our baseline, which was determined by converting the reported magnitude of the star to an expected flux, agrees with the average of the sector but not the individual halves. Bottom: A zoom-in of the middle panel around the 16-hour gray shaded region corresponding to the flyby of (4628) Laplace. This time, we add a constant offset to our model for visual clarity but otherwise do not scale the magnitude of the deviations. We see two separated ``dips'' in the target flux, corresponding to the two times (4628) Laplace crossed a cluster of pixels in the background aperture. Although the target star (likely) did not actually grow fainter during this time, it appeared so relative to the temporarily enhanced background. We do not see a major positive excursion since (4628) Laplace's trajectory never brings the core of its PSF near the red extraction aperture. The small positive excursion corresponds to when the edges of the PSF spill into the red hatched pixel.}
    \label{fig:postprocess_procedure}
\end{figure*}

\begin{enumerate}
    \item For a given target, use bilinear interpolation between the four nearest camera/CCD specific models of the TESS Pixel Response Function (PRF) to generate a function that, when given a nearby sub-pixel position, returns a model of the local PRF\footnote{The files can be accessed \href{https://archive.stsci.edu/missions/tess/models/prf_fitsfiles/}{here}. Our procedures for manipulating these files relies in part on Keaton Bell's open-source \href{https://github.com/keatonb/TESS_PRF}{TESS\_PRF} package and we thank him for making this work available to the community.}. We chose to use model PRFs rather than empirical approximations \citep{anderson_2000, han_2023} for simplicity.
    \item For a given minor planet that may have interacted with this target's TPF, use the World Coordinate System (WCS) pointing information in the Target Pixel File's header to convert the particle's precomputed ephemeris to a time series of TPF pixel coordinates. Using the location-specific model for the PRF, convert this series of pixel positions into a model image cube of the minor planet's motion through the region near the TPF.
    \item Scale each frame of the image cube to match the particle's time-dependent $T$ magnitude as seen from TESS. This is done by transforming the $H$ magnitude reported by the Minor Planet Center\footnote{Specifically, the $H$ magnitudes are pulled from \jorbit's cached MPCORB.DAT file, which was retrieved from the MPC in February 2025.} into a time-dependent $V$ magnitude via the framework of \citet{bowell_1989} and the estimates of the particle's barycentric position and phase angle. Convert the $V$ magnitude to a $T$ magnitude by combining the color transformation in \citet{stassun_2018} and an assumed $J-H$ color of 0.322, the average of common S-type asteroids from \citet{popescu_2018}, which together imply a color of $T-V$ = -0.68\,mag.
    \item Extract the pixel masks used by SPOC when computing the target and background fluxes from the TPF metadata. For every frame in the model cubes, record the magnitude in each aperture from this specific minor planet.
    \item Repeat steps 2-4 for every associated minor planet, then sum the individual time series into total target and background aperture time series.
    \item Repeat the full procedure for every target.
\end{enumerate}

For each target, we were then left with two time series of fluxes from known minor planets: one in the target aperture, and one in the background aperture. These each account for numerous target- and sector-specific effects, including the detector-location specific PRF, the evolving sub-pixel position of the minor planet, the time-varying magnitude of the minor planet caused by changing distances/phase angles, and the target-specific TPF and aperture sizes/positions. These last quantities, the exact size and placement of the apertures, are the most critical for tying these simulations to actual data. Even if a minor planet approaches near a target, and indeed within its TPF, if it contributes little flux to the specific pixels that light curves are built from, it will pass through without impacting the derived time series.

Finally, to compare the model time series with actual data, we created model light curves that mimic SPOC's Simple Aperture Photometry by converting magnitudes to expected count rates via the formula from the TESS Instrument Handbook\footnote{More specifically, we used $\text{T}_{\text{mag}} = -2.5\log_{10}(f) + 20.44$, where $f$ is the flux in e$^-$/s, following the \href{https://tess.mit.edu/public/tesstransients/pages/readme.html}{TESS Transients project} and \citet{takacs_2025}.}, then subtracting the per-pixel background from the sum of all flux in the target aperture.

\subsection{Postprocessing caveats} \label{sub:caveats}
\begin{figure}
    \centering
    \includegraphics[width=\columnwidth]{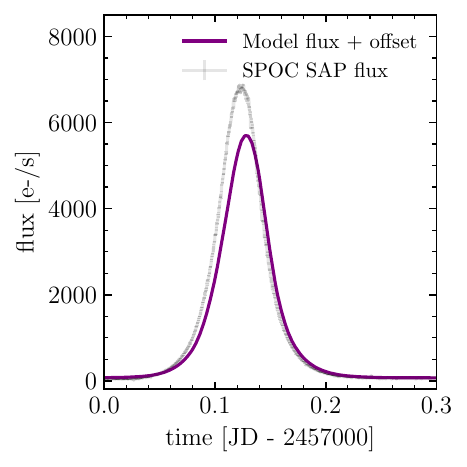}
    \caption{A poor postdiction for the interaction between TIC 388814538 and asteroid (59) Elpis in Sector 43. The offset between peak times is likely the result of slight WCS inaccuracies, while the offset in peak heights could be due to WCS inaccuracies shifting the centroid of the model PSF or inaccurate model flux predictions (see Sec. \ref{sub:caveats} and Appendix \ref{appendix:mag_errs}).}
    \label{fig:bad_example}
\end{figure}

Though our modeling choices in Sec. \ref{sub:postprocessing} generally resulted in excellent agreement with observed SAP data, it is worth highlighting two weakness of our procedure. The first of these is placing absolute trust in the WCS of each TPF header. Manual inspection of a handful of our $>$10,000 model light curves revealed several examples which contained peaks that were offset from the observed SAP flux. These examples also often over or underestimated the amplitude of the minor planet interaction. Fig. \ref{fig:bad_example} shows a representative of this group: even after taking care to undo the light travel time correction that is applied to SPOC timestamps, the peak of the minor planet's contribution is shifted by several minutes from the one observed in the data.

This mismatch in timing and amplitude is due to a sub-pixel error in the header WCS and its resulting effect on the mapping from an on-sky to an on-detector trajectory. Since we generate model images using the sub-pixel position of the asteroid at each timestep, slight shifts in the detector position will cause the asteroid to appear early/late, and possibly closer/further, from the extraction or background apertures. Since these errors do not affect the general presence of an interaction, however, and since when considering the whole population of targets they can expect to balance out, we did not make any effort to correct each individual WCS.

Additionally, the amplitude of each interaction depends on our specific choice of J-H color for the minor planet and the resulting conversion from its $V$\, mag to its magnitude in the TESS band. For a fuller discussion on the impact of these choices and empirical tests of our magnitude postdictions, see App. \ref{appendix:mag_errs}.

\section{General Results} \label{sec:general_results}

\begin{figure*}
    \centering
    \includegraphics[height=0.9\textheight]{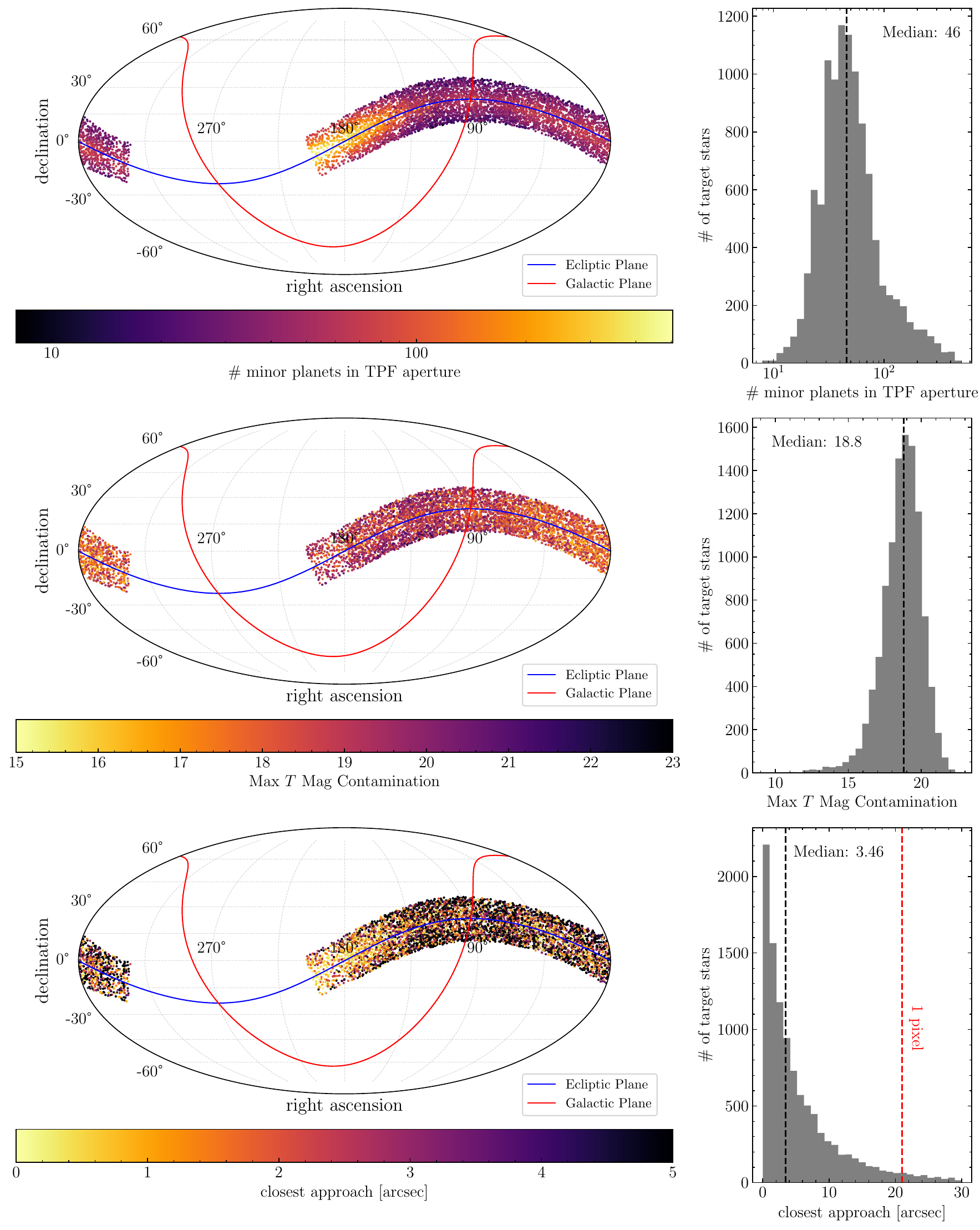}
    \caption{Top: the number of minor planets that interacted with either the target or background extraction apertures on each target-specific TPF. The apparent overdensity near RA=$180^\circ$ is an observational effect explained in Appendix \ref{appendix:overdensity}. Middle: The brightest cumulative, instantaneous $T$ mag from all minor planets crossing the TPF recorded in either the target or background aperture. Bottom: The smallest distance between the target and a minor planet during a given sector. Note that the colorbar cuts off at 5\arcsec although the distribution continues out further, as seen by the subplot to the right.}
    \label{fig:composite_skies}
\end{figure*}

Here, and in Fig. \ref{fig:composite_skies}, we provide overarching population-level results of our simulations. We highlight three main conclusions:

\begin{itemize}
    \item Without exception, every high-cadence target observed during ecliptic plane sectors was affected by minor planets. In most cases, tens of minor planets cross either the background or target extraction apertures.
    \item The median magnitude of all minor planets in one of the apertures at each target's moment of most extreme contamination is 18.8. About 10\% of all target stars experience a moment of $T<17$ mag contamination, meaning a non-negligible fraction of targets experience contamination at the $\sim1\%$ level. As shown in Fig. \ref{fig:contamination_ratios}, about 50\% of all target stars with $13<T<15$ hit this 1\% level, with the percentage rising alongside target magnitude from there.
    \item $>95\%$ of all target stars experience a minor planet crossing within 1 TESS pixel width of the target during a sector's length of observations.
\end{itemize}

\newpage
\newpage

\subsection{Moments of maximum contamination} \label{sub:max_contamination}

\begin{figure*}[!h]
    \centering
    \includegraphics[width=0.8\textwidth]{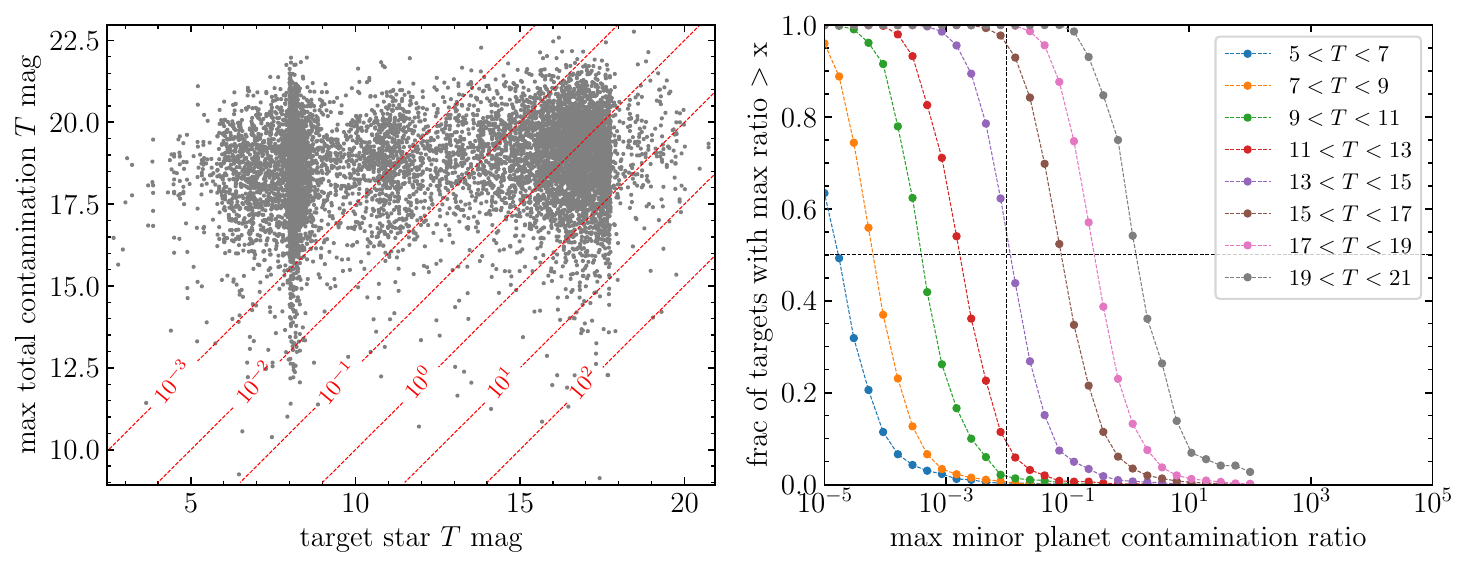}
    \caption{Left: The maximum instantaneous magnitude of the sum of all minor planets within TPF apertures compared to their target TIC's magnitude. Note that the contours of constant flux ratio, thanks to the logarithmic definition of magnitude, are logarithmically spaced: faint targets can be overwhelmed by minor planets, (e.g. the $T=17.4$ TIC 610941254, which in Sector 70, was passed by the $T=9.0$\,mag minor planet (29) Amphitrite, the lowest point on the panel), while bright targets are comparatively unaffected. The bimodality of target $T$\,mags is simply a product of which target stars were accepted as 20\,s targets, mostly from General Investigator proposals. Right: The same data as left, but now binned and shown as a cumulative distribution. A vertical dotted line corresponding to 1\% contamination (minor planet flux/target flux) and a horizontal dotted line corresponding to 50\% of targets within a bin are included as visual aids. Note that this definition of contamination is the brightest moment in either the extraction or background aperture, and that although it was computed by considering contributions from all minor planets, the brightest one typically dominates all others.}
    \label{fig:contamination_ratios}
\end{figure*}

We further investigate the results sketched in the middle panel of Fig. \ref{fig:composite_skies}, the moment of maximum minor planet contamination for each target. We define ``contamination'' as the summed magnitudes of all minor planets in either the target or background apertures for each target. However, we note that typical reduction routines (including our mock SAP light curves like those shown in Fig. \ref{fig:postprocess_procedure}) will likely dilute sources found only in the background aperture since these pixels are only used to determine the mean per-pixel background level. The results are shown in Fig. \ref{fig:contamination_ratios}: as expected, though the distribution of maximum contamination magnitudes is smooth, the actual impact at the population level also depends on the more complex underlying distribution of target magnitudes.

In the left panel of Fig. \ref{fig:contamination_ratios}, we plot the target $T$ mag against the maximum minor planet contamination $T$ mag, along with contours of constant flux ratios as a visual guide. It is clear that even the brightest outlier contamination moments will only meaningfully affect their TPFs if they happen to be associated with relatively faint targets. The right panel of Fig. \ref{fig:contamination_ratios} quantifies this intuition: here we bin the targets by magnitude and compute the fraction of targets in each bin that experience contamination above a range of thresholds. For targets with $T<11$\,mag, we see that vanishingly few experience contamination levels above 1\%. However, in the $13<T<15$ bin, 50\% of targets experience at least 1\% contamination, and by the $15<T<17$ bin, $>90\%$ of targets experience this level of contamination.

\subsection{Serendipitous occultations} \label{sub:occultation_odds}

Considering Fig. \ref{fig:composite_skies}'s illustration that $>95\%$ of high-cadence TPFs of ecliptic targets include moments where a minor planet passes within one pixel of the target star, it is natural to wonder whether serendipitous stellar occultations exist within the TESS archives. These events, defined by moments when a foreground SSO entirely blocks a background star, can enable powerful measurements of an asteroid's shape and size \citep{durech_2011} and provide exquisitely precise astrometry potentially capable of constraining underlying dynamical models of the solar system \citep{rice_2019, gomes_2025}. Unfortunately, however, although two space-based observatories have now observed a minor planet occultation (CHEOPS, \citealt{morgado_2022}; JWST, \citealt{santos_sanz_2017}), we show here that TESS is unlikely to join their ranks, and that even if it does, the science returns will be severely limited.

The primary challenge confronting TESS is the formidable timing resolution required to resolve most occultations. Main belt asteroids, by far the most common type of SSO and therefore the most likely to be involved in a serendipitous occultation, typically move at on-sky speeds that produce $\sim$1\,s stellar crossing times \citep{van_Altena_2012}. Though this is plausibly detectable with 20\,s TESS observations (i.e. a single data point might appear $\sim5\%$ below the mean should 100\% of the flux be blocked 1/20th of the exposure), the extreme ``smearing'' and flux dilution of the event significantly limits both the science benefits and the detection potential.

More unfortunately, one cannot fall back on the typical argument deployed when considering many of TESS's limitations, that its enormous field of view encompasses so many targets that rare events still stand a solid chance of detection. This is because all other data products besides the 20\,s TPFs, meaning all longer-cadence TPFs and the FFIs, employ an onboard Cosmic Ray Mitigation strategy\footnote{See \href{https://heasarc.gsfc.nasa.gov/docs/tess/TESS-CosmicRayPrimer.html}{this primer} from the TESS General Investigator Office.} that discards the brightest and faintest 2\,s integration in each 20\,s batch before stacking. This means any $<2$\,s events outside of the 20\,s TPFs will likely never be downlinked.

Even still, we can use our postdiction simulations to further investigate the likelihood of a serendipitous occultation. We begin this effort by noting that everything which follows are simply point estimates that use the best-fit orbit for each minor planet and do not consider the uncertainties in their predicted ephemerides. In other words, it is possible that some minor planets predicted to ``miss''  their targets do in fact undergo an occultation, while those predicted to ``hit'' will, in fact, pass wide. Additionally, we use a simplified occultation model that does not account for diffraction or smearing effects \citep{roques_2000}. We leave a more rigorous treatment that uses a more realistic model and marginalizes over the complex time-dependent uncertainties in both the minor planet and spacecraft positions to others who wish to actually search for these single-exposure ``dips'': we did not undertake a systematic search as a part of this study.

We first set aside every minor planet that passed within 5\arcsec\, of a target in our \jorbit simulations and used the ``Stellar Occultation Reduction and Analysis'' package  \citep[\texttt{SORA},][]{sora} to compute specific occultation parameters for each minor planet/target combination. Of the total of 13,313 unique minor planet-target pairs from \jorbit, \texttt{SORA} ran successfully for 12,976; most failed attempts crashed due to unsuccessful database matching (SORA relies on the SIMBAD database \citep{simbad} to check for proper motion and other stellar properties), though one stemmed from updates to a particle's best-fit orbit between \jorbit's cache creation and Horizons' present-day estimate\footnote{2021 VB39, a minor planet that \jorbit predicted flew within 3\arcsec of TIC 114058447 in Sector 42, was included in the MPC's Daily Orbit Update on Feb. 14, 2025 (MPEC 2025-C169), a few weeks after \jorbit's cache was created. The new best-fit orbit shifted 2021 VB39's initial position on 2020-01-01 by 0.033\,AU and its initial velocity by 0.11\,m/s, a change large enough to push the predicted close approach time outside of \texttt{SORA}'s search window.}.

\texttt{SORA} retrieves information on TESS's orbital position, the trajectory of each minor planet, and the proper motion/parallax of each target star to compute the instantaneous shadow velocity, distance to the minor planet, angular separation, and time for the moment of closest approach between each pair. To convert the shadow velocity to a conservative star-crossing timescale (under the assumption that a) an occultation actually occurs and that b) the center of the asteroid passes perfectly over the center of the star), we convert each minor planet's absolute $H$ magnitude into a diameter via a simple scaling \citep{harris_1997} and an assumed albedo of 0.1. Note that this heuristic choice of albedo sets the implied asteroid diameters and therefore affects the odds of an occultation, meaning the results that follow should be regarded as a weak function of albedo.

\begin{figure}
    \centering
    \includegraphics[width=\columnwidth]{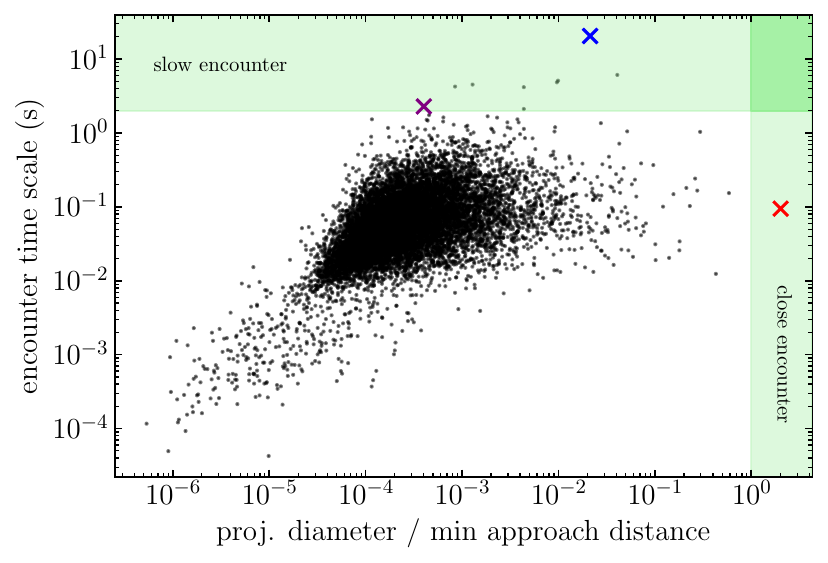}
    \caption{The predicted closest approach distance (normalized by minor planet projected diameter) and the timescale of each encounter for all flybys with a minimum distance of $5$\arcsec. The horizontal green shaded region marks all interactions which last $>2$\,s, meaning they would be preserved in data products that employ onboard cosmic ray mitigation (everything aside from the 20\,s TPFs). Note, however, that encounters outside of this band are potentially detectable in the 20\,s data (see Fig. \ref{fig:occultation_snr}). The vertical band illustrates the region where \texttt{SORA}, using the best-fit orbit for each minor planet, actually predicts an occultation will occur. No predicted interactions fall in the union of these two regions, though uncertainties in TESS's trajectory and minor planet orbits could in principle conspire to produce an occultation. The highest SNR encounter is marked in purple, the longest encounter in blue, and the closest encounter in red. See Sec. \ref{sub:occultation_odds} for details.}
    \label{fig:occultation_timescale}
\end{figure}

We use these diameters and the instantaneous distances to each minor planet to compute each sky-projected angular diameter, and compare this to the angular distance of closest approach. These results are shown in Fig. \ref{fig:occultation_timescale}, which reveals that the closest predicted approach is often many 100s of times greater than the projected diameter, and that even if the ephemeris uncertainty produced an occultation despite the predicted miss, the majority of the encounters will be $<1$\,s long.

\begin{figure}
    \centering
    \includegraphics[width=\columnwidth]{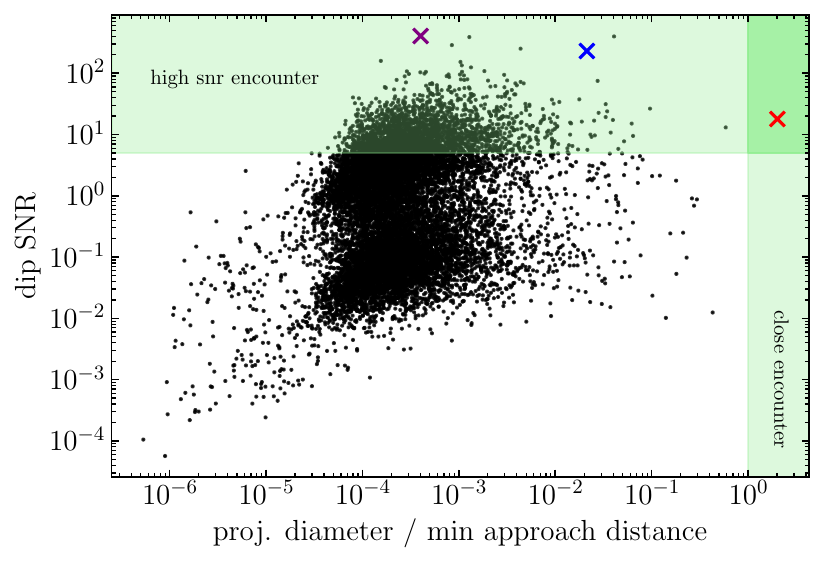}
    \caption{Similar to Fig. \ref{fig:occultation_timescale}, except that instead of physical timescale, we show the expected SNR of dip in flux caused by an occultation, even if that occultation is shorter than a 2\,s frame. The bimodal distribution in expected SNRs follows from the bimodal distribution in target magnitudes also seen in Fig. \ref{fig:contamination_ratios}; a sufficiently bright target star, if blocked for even a short time, can reveal an occultation with high statistical confidence thanks to the large number of photons intercepted by the minor planet. Like Fig. \ref{fig:occultation_timescale}, the highest SNR encounter is marked in purple, the longest encounter in blue, and the closest encounter in red.}
    \label{fig:occultation_snr}
\end{figure}

As discussed previously, however, an event shorter than the exposure time is still potentially detectable, even if precise shape/timing information is unrecoverable. For an alternative view of the same results presented in Fig. \ref{fig:occultation_timescale}, we consider not the physical timescale, but the estimated signal to noise ratio (SNR) of the drop in flux for the single exposure that contains the occultation. To do this, we estimated the raw number of counts received on the detector during a 20\,s exposure using the same $T$\,mag to counts conversion used in Sec. \ref{sub:postprocessing}. We then assumed that this was the mean signal level and could be estimated by considering a baseline much larger than a single exposure, which allows us to consider only the size of the ``dip'', not the comparison between the dip and the mean. We compute the number of counts lost due to each interaction-specific occultation, then assume perfectly Poisson noise given the mean level to arrive at an estimated SNR. These results are shown in Fig. \ref{fig:occultation_snr}.

In addition to Figures \ref{fig:occultation_timescale} and \ref{fig:occultation_snr}, we also describe a subset of the most extreme predictions. The largest ratio between projected asteroid diameter and closest approach distance (indeed, the only pair with a ratio $>1$, implying that the best-fit orbit predicts an actual occultation) involves TIC 363549733 in Sector 45 and the minor planet (85566) 1998 BM9. This encounter is shown in red in Figs. \ref{fig:occultation_snr} and \ref{fig:occultation_timescale}. In our scheme, this object has a diameter of 3.66\,km, a projected diameter of 1.49\,mas, and a minimum approach distance of 0.42\,mas. Moving at 0.0092\arcsec/s, the encounter would have lasted at most 0.16\,s. The longest potential encounter could have taken place between TIC 400465017 in Sector 72 and the minor planet (704) Interamnia. Though the best-fit orbit placed the asteroid 4.3\arcsec away, should the 159\,mas object have deviated (unreasonably) far from its predicted trajectory and occulted the target star, the encounter would have lasted 35.6\,s. This encounter is shown in blue in Figs. \ref{fig:occultation_snr} and \ref{fig:occultation_timescale}. Finally, the highest SNR encounter could have taken place between TIC 376983076 in Sector 70 and the Trans-Neptunian Object 2013 ST102. This distant object would have taken 2.2\,s to cross this extremely bright target star ($T$=5.94), a combination that would have fully blocked the otherwise saturating flux level for a full 2\,s frame and produced a dip with an SNR of 406. This encounter is shown in purple in Figs. \ref{fig:occultation_snr} and \ref{fig:occultation_timescale}. We checked the real TESS data from the SPOC pipeline for each of these targets and found no evidence of any occultations.

Overall, Fig. \ref{fig:occultation_snr} offers a comparatively optimistic view when held up to Fig. \ref{fig:occultation_timescale}: when considering bright, high-SNR targets, even rapid occultations will likely leave a statistically significant mark on a single exposure. However, again, since such an occultation would not extend beyond a single data point with the current or past TESS survey strategy, we did not embark on a systematic search for minor planet occultations. We do emphasize that TESS is in principle capable of detecting such events, and could even target predicted longer-duration occulations involving more distant SSOs in future cycles.

\subsection{Characteristic Timescales} \label{sub:timescales}

Finally, the last quantity we consider at the population level is the typical timescale of a minor planet's interaction with a high-cadence TPF. Note that we frame this as an ``interaction'' timescale, rather than a peak width or brightening timescale: some interactions, like that in Fig. \ref{fig:postprocess_procedure}, contain both positive and negative deviations, while others, like those illustrated in Sec. \ref{sec:case_studies}, only produce positive excursions from the baseline target flux. The shape of each specific interaction is governed by the particular minor planet's trajectory through the particular TPF's selection of target and background apertures, meaning one should not assume that in general a minor planet flyby always leads to an apparent brightening.

\begin{figure}
    \centering
    \includegraphics[width=\columnwidth]{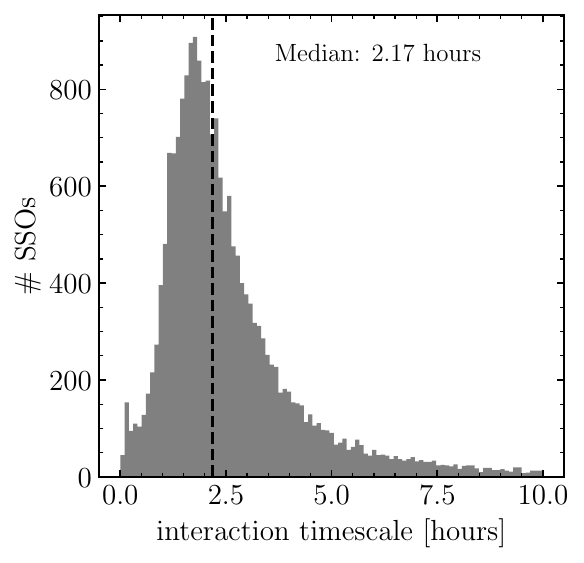}
    \caption{The interaction timescale between minor planets and target stars from our model SAP time series. Note that the median timescale is comparable to the transit duration of a short-period planet, and that we have excluded all peaks with a width of $<15$\,min since we consider these to be largely spurious and driven by numerical issues of very small peaks caused by the edges of minor planet PSFs leaking into extraction apertures.}
    \label{fig:peak_timescales}
\end{figure}

To measure this timescale, we applied a simple peak-finding routine from the \texttt{scipy} package \citep{scipy} to the absolute value of each of our model SAP time series. This extracts all deviations, positive and negative, meaning that one minor planet flyby may be associated with multiple interactions (again, see Fig. \ref{fig:postprocess_procedure} for an example of alternating positive/negative excursions, and Fig. \ref{fig:case_studies} for examples of one interaction/flyby). The distribution of all peaks is shown in Fig. \ref{fig:peak_timescales}, which illustrates that the typical minor planet-induced deviation lasts on the order of a few hours, though there is a long tail towards longer durations. This timescale is comparable to the transit duration of short-period exoplanets, exactly the type of object TESS is optimized to discover, which implies that a minor planet flyby could in principle mimic a planetary transit (e.g., via inducing a dip like that shown in Fig. \ref{fig:postprocess_procedure}).

Thankfully minor planet flybys are unlikely to produce periodic dips that most survey-level planet detection algorithms rely on  \citep{hippke_2019}. However, so-called ``mono-transits'', or isolated transits without an observed repeat, are of considerable interest to the community as they could potentially reveal longer-period planets and probe a challenging-to-explore demographic region \citep{osborn_2016, magliano_2024}. We urge the mono-transit community to use tools like \jorbit to rule out minor planet contamination as a possible source of false positives.

\section{Case Studies} \label{sec:case_studies}

\begin{figure*}
    \centering
    \includegraphics[width=0.7\textwidth]{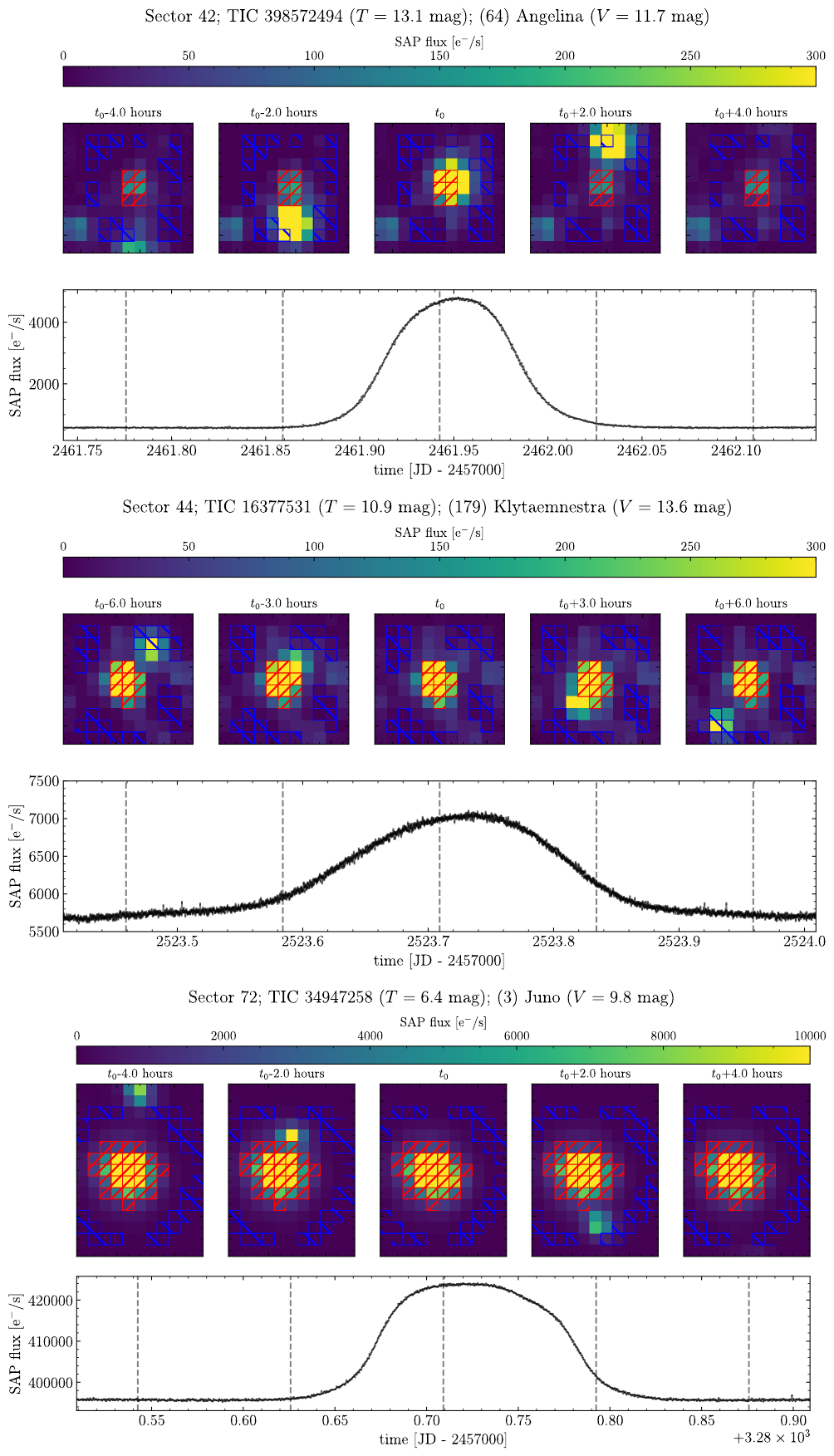}
    \caption{Three examples of encounters between high-cadence targets and bright asteroids. The images and lightcurves are real data from the SPOC pipeline, not simulations. The vertical lines on the time series mark the times of the image snapshots. See Sec. \ref{sec:case_studies} for details.}
    \label{fig:case_studies}
\end{figure*}

While above we discussed population-level results derived from the entire collection of targets and minor planet encounters, here we zoom in and consider several dramatic individuals of that population. Fig. \ref{fig:case_studies} shows three target/minor planet encounters that are among the brightest predicted by our simulations, and shows that all three were indeed recovered cleanly in the archived data.

In this figure, the images are real slices from the TPFs (retrieved via \texttt{lightkurve} plotted with the same color bar in each row), while the flux is the Simple Aperture Photometry flux from the corresponding light curve. Following the same pattern as Fig. \ref{fig:postprocess_procedure}, the red hatched squares mark the target extraction aperture, while the blue hatched squares mark the pixels composing the background aperture. We use $t_0$ to denote the predicted moment of minimum separation between the minor planet and the target star.  Note, however, that this does not always correspond exactly to the moment of maximum contamination as this depends on the sub-pixel position/PRF shape and the layout of the aperture pixels.

Note that, while Fig. \ref{fig:postprocess_procedure} demonstrated how minor planets can produce apparent dimming events by crossing the background aperture, in these cases the target star appears to grow brighter as the minor planet crosses the extraction aperture. This highlights the variable morphology of asteroid/star interactions: not every encounter produces the dip-spike-dip pattern seen in Fig. \ref{fig:postprocess_procedure}, and the exact scale/sign of the deviation will be unique to the flyby geometry. In principle, the former type of interaction could mimic an exoplanet transit, while this latter type could mimic astrophysical processes that produce temporary brightenings such as stellar flares or microlensing events.

The first of these cases, for TIC 398572494 and the asteroid (64) Angelina, provides an example of a star being overwhelmed by a minor planet. Over a roughly three hour window, the flux increases nearly 500\% before returning to baseline again. The second case, TIC 16377531 and the asteroid (179) Klytaemnestra, similarly shows a dramatic temporary increase in flux. The final case, TIC 34947258 and the asteroid (3) Juno, is particularly interesting. Juno is one of the largest/brightest asteroids in the solar system, but although in terms of raw flux it had the largest contribution to its light curve of these examples, it actually had the smallest fractional effect. That is because TIC 34947258 is an extremely bright star for a TESS target at $T=6.5$\,mag, and actually saturates a portion of the detector. As a result, this TPF is much larger than the standard 11x11 pixels, and the extraction aperture is also larger than normal.

\section{Conclusion} \label{sec:conclusion}

To summarize, we have created model light curves of the contributions from solar system objects for every 20\,s cadence TESS target in pre-2025 ecliptic plane sectors. These models are built from \jorbit N-body simulations of every known minor planet that passed near enough to each specific target to affect its SPOC Simple Aperture Photometry. The conversion from an astrometric trajectory to a light curve relies on a realistic PRF model, the actual target and background apertures assigned to each TPF by SPOC, and calculations of the time-varying apparent magnitude of each minor planet based on data curated by the Minor Planet Center.

At the population level, we find that minor planets flew through every target TPF, and that in $>95\%$ of cases, one or more approached within 1 pixel of the target. The median of the instantaneous maximum contribution from minor planets across all targets is $T$\,mag of 18.8, and $>50\%$ of all targets with $T>13$\,mag experience at least one moment in a sector's worth of observations where the flux from background minor planets is at least 1\% of the target flux. Minor planets can cause either apparent brightenings or dimmings in an SAP light curve depending on their trajectory relative to the extraction and background apertures. The median timescale of a minor planet's interaction with a TPF is 2.17 hours.

We also presented four individual case studies in Fig. \ref{fig:case_studies} which collectively illustrate the dramatic range of effects minor planets can have on high-cadence data. While Fig. \ref{fig:postprocess_procedure} demonstrates that minor planet flybys can lead to spurious apparent dimming events, the examples in Fig. \ref{fig:case_studies} demonstrate that they can also cause apparent brightenings. Though their amplitude, morphology, and timescale will vary interaction to interaction, minor planets can in principle crudely mimic a wide range of astrophysical phenomena, including exoplanet transits, stellar flares, and microlensing events.

Collectively, these simulations quantify the impact of minor planets on all data recorded in ecliptic plane sectors. The implications of these impacts vary depending on the intended use of that data, and they are more or less important for different science cases. For programs that rely on measurements of only bright targets and can tolerate sub-percent biases, minor planets do not meaningfully contribute to the light curves. However, for programs designed to measure subtle variations of faint targets, it is possible for a signal of interest to be masked or even mimicked by a passing minor planet.

Thankfully these events can be accurately pre or postdicted by propagating an appropriate minor planet's orbit to the times of observation using tools like \jorbit, which can also help identify the appropriate minor planets to check in the first place. We hope that this study assures the community that minor planet contamination can be accounted for and serves as a template for how to check specific observations for contributions from foreground minor planets.

{\begin{acknowledgments}

D.G., E.M., and A.R. were supported by the Student Training in Astronomy Research (STAR) program at Columbia University, which is grateful for the support of the Pinkerton Foundation, New York Science Research Mentoring Consortium, and the National Osterbrock Leadership Program of the AAS. D.A.Y. thanks the LSST-DA Data Science Fellowship Program, which is funded by LSST-DA, the Brinson Foundation, and the Moore Foundation; his participation in the program has benefited this work.
This work was supported in part by NASA TESS GI grant \#80NSSC24K0359 and Heising-Simons Foundation Grant \#2023-4478. This research has made use of the Astrophysics Data System, funded by NASA under Cooperative Agreement 80NSSC21M00561.
Data from the MPC's database is made freely available to the public. Funding for this data and the MPC's operations comes from a NASA PDCO grant (80NSSC22M0024), administered via a University of Maryland - SAO subaward (106075-Z6415201). The MPC's computing equipment is funded in part by the above award, and in part by funding from the Tamkin Foundation.
We thank the Yale Center for Research Computing for guidance and assistance in using the Grace cluster.
B.C. thanks Diandra Cassese for her patient tutorials on how to use Inkscape to create composite figures.

\end{acknowledgments}}

\vspace{5mm}
\facilities{TESS, Grace high-performance computing cluster at Yale University}

\software{
    \jorbit \citep{jorbit};
    \texttt{astropy} \citep{astropy:2013, astropy:2018, astropy:2022}; \texttt{SORA} \citep{sora};
    \texttt{lightkurve} \citep{lightkurve};
    \texttt{matplotlib} \citep{hunter2007matplotlib};
    \texttt{numpy} \citep{oliphant2006guide, walt2011numpy, harris2020array}; 
    \texttt{scipy} \citep{scipy}
}

\appendix
\section{Apparent Over-density of Minor Planets at RA=\texorpdfstring{$180^\circ$}{180 degrees}} \label{appendix:overdensity}

\begin{wrapfigure}{l}{0.5\textwidth}
    \centering
    \includegraphics[width=0.48\textwidth]{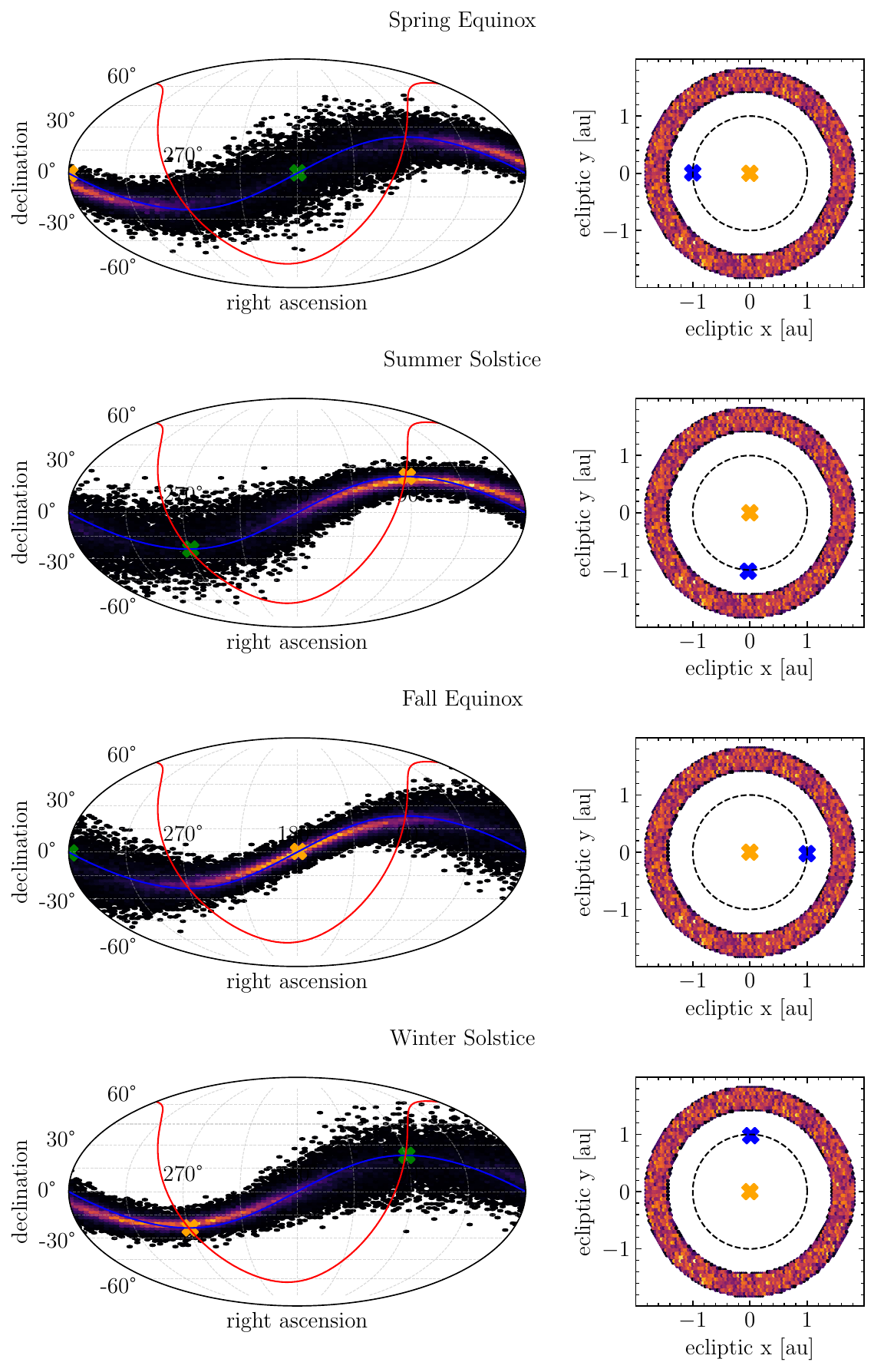}
    \caption{A toy model of the solar system to illustrate how the Earth's off-center position can produce an apparent anisotropic distribution of minor planets whose peak moves throughout the year. The orange "x"s mark the position of the Sun, as seen on sky in the left columns and in 3D space in the right columns. The green "x" marks the anti-solar position on sky, and the blue "x" marks the position of Earth. The ecliptic plane is marked in blue, while the galactic plane is marked in red. The color scaling is arbitrary and based on the model described in the text.}
    \label{fig:overdensity_toy_model}
\end{wrapfigure}
In the top panel of Fig. \ref{fig:composite_skies}, we see that the number of minor planets that interact with the 20\,s cadence TPFs is not uniform in ecliptic longitude. This is perhaps somewhat surprising, since the distribution of minor planets is largely isotropic in longitude. Indeed, the apparent over-density near RA=$180^\circ$ is the final product of three competing observational effects induced both by TESS's time-dependent position within the solar system and its particular pointing strategy. We dedicate this Appendix to illustrating these effects.

We begin not with the real solar system, but rather, a toy model. We randomly place a large number of ``asteroids'' with $2.1<r<3.2$\,au, ecliptic longitude $\phi$ uniform between $0^\circ$ and $360^\circ$, and ecliptic latitude $a$ set via a sharp beta distribution with $a=1, b=20$ to mimic the asteroid belt (the choice of latitude distribution is unimportant and unconnected to the real solar system; these values were chosen heuristically to give the belt a finite width). We then place the Earth at several notable positions throughout the year, and \textit{without regard for position-dependent magnitudes}, compute the density of asteroids seen on-sky (without accounting for light travel time corrections). The sky maps and ``top-down'' views of the solar system are shown in Fig. \ref{fig:overdensity_toy_model}.

Although the ring of asteroids is uniform, our view of them is not, since the Earth is not at the center of the solar system.  A given field of view, when aimed (impractically) directly at the Sun, will subtend a much larger portion of the asteroid belt than if it were aimed directly \textit{away} from the Sun. Consequently the on-sky number density of asteroids has a ``hot spot'' whose location shifts throughout the year, tracking the Sun.

\begin{figure}
    \centering
    \includegraphics[width=\textwidth]{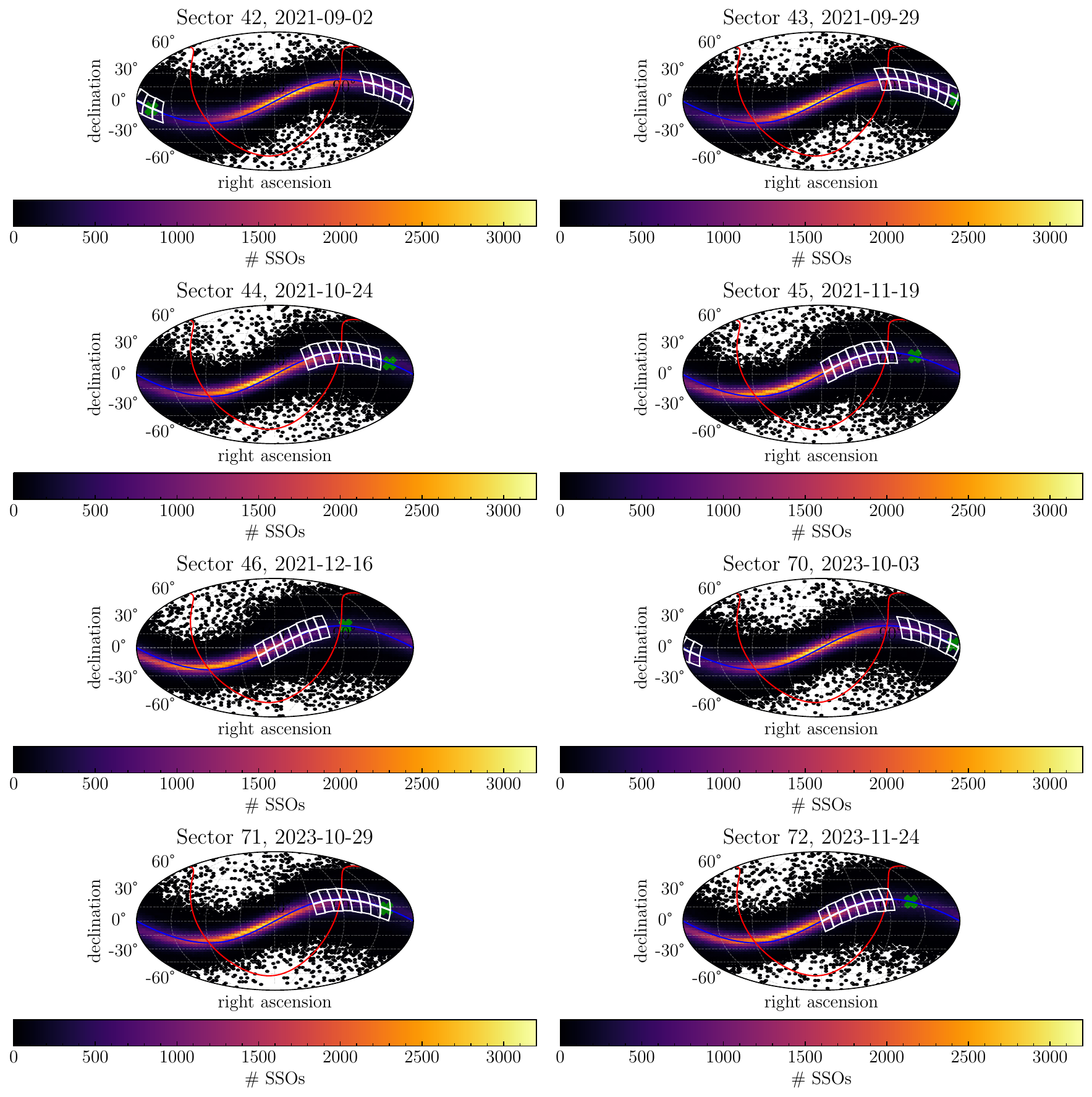}
    \caption{The actual distribution of minor planets shown at the time of each individual ecliptic plane sector. Note that TESS does not center its ecliptic observations on the mean anti-solar point (marked with a green ``x'' in each subplot), as it does in its standard vertically-oriented observing mode. In all subplots, the ecliptic plane is marked in blue, while the galactic plane is marked in red.}
    \label{fig:overdensity_tess}
\end{figure}

Fig. \ref{fig:overdensity_tess} displays a more realistic view of this effect. For every ecliptic-plane sector, we used \jorbit to compute the sky positions of all $\sim1.4$\,million known real minor planets at the mid-time of the sector. We also plot the footprint of each of TESS's 16 CCDs, and the position of the anti-solar point at the midpoint. Note that 1) the same effect simulated by the toy model is still present, and 2) the overall TESS footprint is not centered on the anti-solar point, as it is when TESS is not observing the ecliptic. Consequently, we are left with a non-symmetric slice out of this over-density, which largely explains the over-density seen in Fig. \ref{fig:composite_skies}.

\begin{figure}
    \centering
    \includegraphics[width=0.75\textwidth]{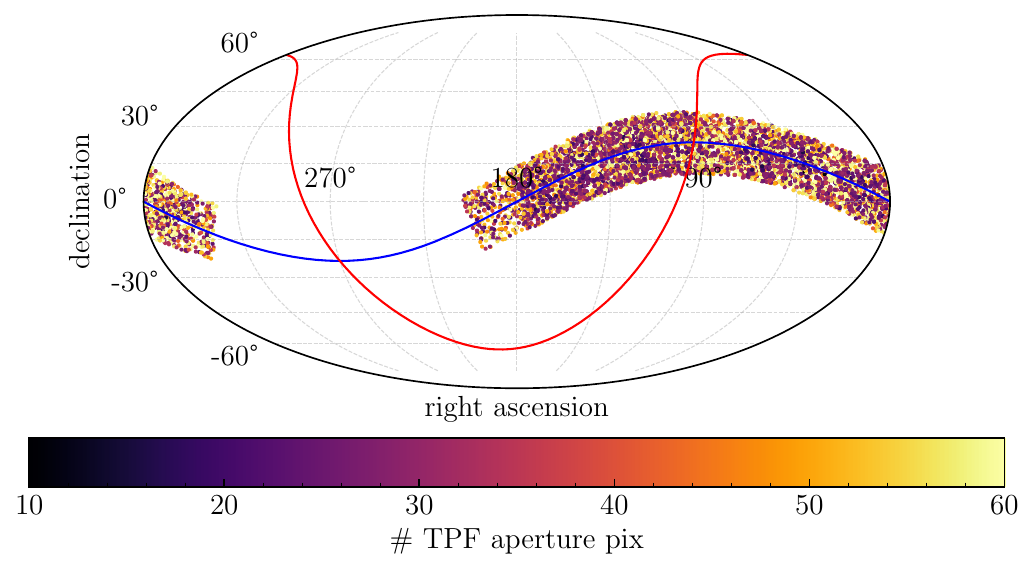}
    \caption{The total number of TESS pixels that make up the extraction and background apertures for each target as seen on sky. While there is a slight bias towards larger apertures farther from the galactic plane where crowding is less of a concern, the pattern is not as strong as the observational bias shown in Fig. \ref{fig:overdensity_tess}. The ecliptic plane is marked in blue, while the galactic plane is marked in red.}
    \label{fig:num_relevant_pix}
\end{figure}
In addition to this effect, there are two other practical consequences of observation to consider. Firstly, objects at opposition appear much brighter than those at conjunction, meaning that there is a competing bias that works to cancel out the viewing effect. Finally, Fig. \ref{fig:composite_skies} displays the number of minor planets that imprint on each target-specific collection of background or target aperture. Since the size of each aperture depends on the target star and its surrounding environment (e.g. bright objects have large target apertures, while those in crowded environments like the galactic plane will have smaller background apertures), the size of each ``net'' to catch minor planets also varies across the sky, though we find in general that it does not do so systematically in Fig. \ref{fig:num_relevant_pix}.

\section{Impact of Magnitude/Color Uncertainties} \label{appendix:mag_errs}

This study used a forward model to convert asteroid orbits into predictions of their flux as observed by TESS over time. For each individual asteroid, this conversion explicitly depends on several assumptions:
\begin{enumerate}
    \item The asteroid's absolute magnitude $H$ is well-constrained, and can be converted to a trustworthy $V$\, mag via the relation in \citet{bowell_1989}.
    \item The asteroid's color is exactly known. We used a value of $J-H=0.322$, which is typical of S and C type asteroids which collectively make up the vast majority of known minor planets \citep{popescu_2018}.
    \item The conversion from a $V$\,mag and a color to TESS $T$\,mag, taken from \citet{stassun_2018} and calibrated on stars, is applicable to asteroids.
\end{enumerate}
There is little that can be done to assuage the first point, as the MPC does not distribute uncertainties on the absolute magnitude of minor planets, and in general the phase function of a minor planet is not constrained at all (leading the MPC to assign a typical value of $g=0.15$). The second and third assumptions, however, deserve further scrutiny.

We used a fiducial value of $J-H=0.322$, but \citet{popescu_2018} also report uncertainties on their average colors: $J-H=0.322 \pm 0.056$ for S type asteroids and $J-H=0.320\pm0.025$ for C type. Since the distribution of viable colors for C type asteroids is entirely contained within the distribution of reasonable colors for S type asteroids, we conservatively use the uncertainty on S type colors to quantify the impact of color uncertainties.

Consider the asteroid (3) Juno, the brightest asteroid shown in Fig. \ref{fig:case_studies}. (3) Juno has an absolute magnitude of $H=5.19$. Assuming we observe it at opposition, the conversion of \citet{bowell_1989} gives us a $V$ magnitude of $V=8.43$\,mag. We drew 100,000 samples from $J-H=0.322\pm0.056$, passed them through the color conversion of \citet{stassun_2018} to get $T-V$ colors, which produced distribution of $T$ mags with $T=7.762\pm0.039$. Passing this distribution through the handbook conversion from $T$ to flux gives a distribution of $f=117829\pm4266$\,e$^{-}/s$, or a width equal to 3\% of the mean. Since mathematically this procedure is actually independent of absolute mag $H$, we can conclude that uncertainty in asteroid colors induces a 3\% uncertainty in each of our asteroid flux estimates.

The third assumption, that the color conversion of \citet{stassun_2018} is applicable to asteroids although it was calibrated on stars, requires a comparison between our predicted fluxes and those actually observed in TESS data to verify. Doing so for every minor planet in our sample is beyond the scope of this study, as is a thorough investigation into optimal light curve extraction for a minor planet. However, we can roughly address this assumption by investigating a subset of minor planets using an ``out-of-the-box'' tool such as the \texttt{tess\_asteroids} package \citep{tess_asteroids}.

We used \jorbit to identify all minor planets that fell within 3 degrees of the center of each CCD on December 10th, 2021, which is several days into Sector 46. We chose this ecliptic-oriented sector as it is centered relatively far from the galactic plane and thus avoids dense background fields, and this date to avoid the initial settling time associated with scattered light at the beginning of each sector. We then down selected to only consider minor planets with absolute magnitudes $H<12$ and predicted fluxes in the TESS $T$ band of $>10$e$^-$/s (as calculated via the $H$ to $T$ conversion described above and in Sec. \ref{sec:simulations}), leaving us with a sample of 88 objects.

\begin{figure}
    \centering
    \includegraphics[width=0.5\textwidth]{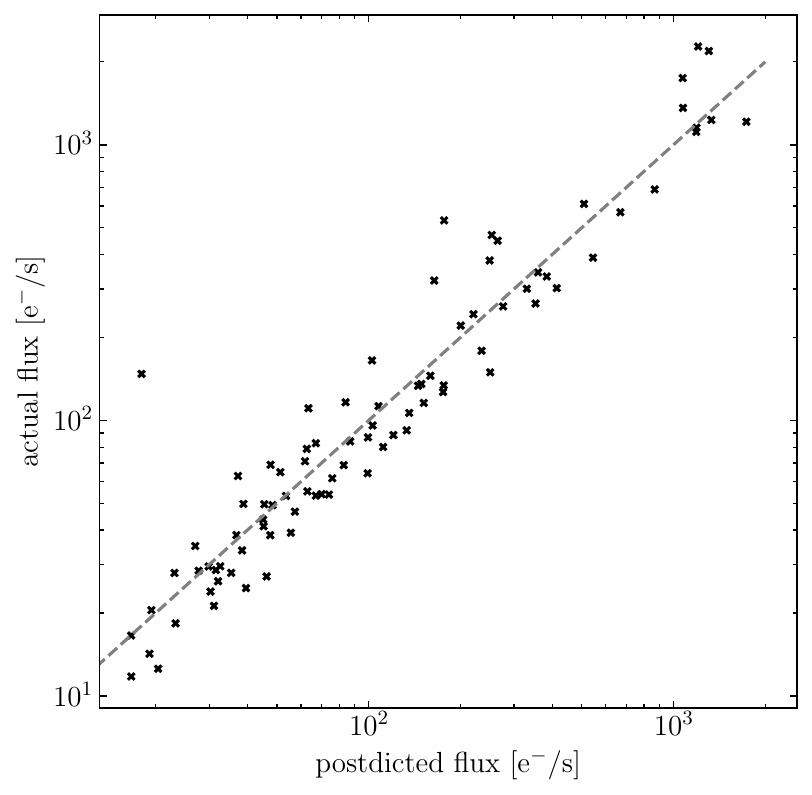}
    \caption{The fluxes postdicted using the procedure described in Sec. \ref{sec:simulations} compared to the fluxes as extracted from TESS data using \texttt{tess\_asteroids} for 88 minor planets selected from Sector 46. The dashed grey line marks 1:1 agreement. These flux values correspond to $T$ mag values between 12 and 17.}
    \label{fig:color_issues}
\end{figure}

We then used the default settings of \texttt{tess\_asteroids} to extract 1-day light curves of each object centered on December 10th and took the mean flux over this time span as an estimate of each minor planet's empirical flux. Finally, we compared the fluxes postdicted with the star-based color conversion from \citet{stassun_2018} to these empirical measurements. This comparison is shown in Fig. \ref{fig:color_issues}, which illustrates that our method produces postdictions that are in good general agreement with the empirical fluxes across a range of minor planet magnitudes.

Though it is impossible to untangle the errors associated with any of the three assumptions described above, as well as any errors introduced by the light curve extraction procedure (e.g. assumptions about background subtraction), we see that all told our flux-prediction scheme well approximates the empirically observed fluxes for this subset of minor planets.

\bibliography{references}{}
\bibliographystyle{aasjournalv7}

\end{document}